\documentclass[aps,twocolumn,superscriptaddress]{revtex4-1}
\usepackage{amsmath} %per le formule matematiche
\usepackage{braket} %per la notazione di dirac
\usepackage{amsfonts} %per l'operatore identità \mathbb{I}
\usepackage{booktabs} %toprule tabelle
\usepackage{graphicx} %importazione immagini
\usepackage{float} %posizionamento H immagini
\usepackage{multirow}
\usepackage[caption=false]{subfig}
\usepackage{xcolor}
\usepackage{physics}
\usepackage{hyperref}
\usepackage{soul}
\begin{document}
\title{Two-qubit quantum probes for the temperature of an Ohmic environment}
\author{Francesca Gebbia}
\affiliation{Department of Physics, University of Trieste, I-34151 Trieste, Italy} 
\author{Claudia Benedetti}
\affiliation{Quantum Technology Lab, Dipartimento di Fisica 
{\em Aldo Pontremoli}, Universit\`a degli Studi di Milano, 
I-20133 Milano, Italia}
\author{ Fabio Benatti}
\affiliation{Department of Physics, University of Trieste, I-34151 Trieste, Italy} 
\author{Roberto Floreanini}
\affiliation{Istituto Nazionale di Fisica Nucleare, Sezione di Trieste, I-34151 Trieste, Italy}
\author{Matteo Bina}
\affiliation{Quantum Technology Lab, Dipartimento di Fisica 
{\em Aldo Pontremoli}, Universit\`a degli Studi di Milano, 
I-20133 Milano, Italia}
\author{Matteo G. A. Paris}
\affiliation{Quantum Technology Lab, Dipartimento di Fisica 
{\em Aldo Pontremoli}, Universit\`a degli Studi di Milano, 
I-20133 Milano, Italia}
\begin{abstract}
We address a particular instance where open quantum systems 
may be used as quantum probes for an emergent property of a complex 
system, as the temperature of a thermal bath. The inherent 
fragility of the quantum probes against decoherence is the key feature 
making the overall scheme very sensitive. The specific setting examined 
here is that of quantum thermometry, which aims to exploits decoherence 
as resource to estimate the temperature of a sample. We focus on 
temperature estimation for a bosonic bath at equilibrium in the Ohmic 
regime (ranging from sub-Ohmic to super-Ohmic), by using pairs of qubits 
in different initial states  and interacting with different environments, consisting either of a single thermal bath, or of two independent ones 
at the same temperature. Our scheme involves pure dephasing of the probes, 
thus avoiding energy exchange with the sample and the consequent perturbation
of temperature itself. We discuss the interplay between correlations among 
the probes and correlations within the bath, and show that entanglement 
improves thermometry at short times whereas, if the interaction time is not constrained, coherence rather than entanglement, is the key resource in 
quantum thermometry.
\end{abstract}
\maketitle
\section{Introduction}
Quantum sensing techniques are among the most
advanced quantum technologies and have led to major changes in 
the field of metrology in the last two decades. Upon exploiting 
the peculiar features of quantum systems, several novel 
enhanced sensors and measuring devices have been indeed 
suggested and demonstrated. In particular, quantum sensing 
based on quantum coherence and entanglement have been exploited 
to overcome precision bounds of classical sensors
\cite{DegenRMP17, lqe09, braun,smirne}. 
Among quantum
sensing techniques, the use of quantum probes has 
recently gained attention as a non-invasive technique to 
estimate parameters of interest without perturbing too much 
the system under investigation. The basic scheme is the following:
a simple quantum system, say a qubit or a pair of qubits, 
is prepared in a given initial state, and then interacts with 
an external system under investigation. After the interaction, which 
imprints information of some parameter onto the state of the quantum 
probe, the latter is measured in order to extract such information
\cite{benedett14,giorgi16,nokkala16,cosco17,usui18}. 
\par
In this paper, we address the use of quantum probes to estimate 
an emergent property of a complex system, i.e. its temperature 
\cite{brunelli11,correa15,giovan17,campbell18,correa18,correa19}.
In particular, in view of its importance for several fields of
quantum information science, we consider here quantum 
thermometry of a bosonic bath in the Ohmic regime, ranging from 
sub-Ohmic to super-Ohmic. At first sight, quantum features may 
not be expected to play a role in building an effective thermometer. 
After all, temperature is an inherently classical parameter and 
any change in the temperature of a sample is just changing its 
(classical) equilibrium distribution. On the other hand, 
quantum probing involves out-of-equilibrium states and since
temperature is governing the amount of thermal fluctuations,  
the inherent fragility of quantum systems against decoherence 
is the key feature making the overall thermometric 
scheme very sensitive. 
\par
The use of single qubit as a thermometer has been analyzed
recently, illustrating the interplay between the dephasing 
dynamics and the Ohmic structure of the environment in 
determining the overall precision \cite{razavian19,razavian2019}. 
Here, we devote attention 
to the role of   correlations in the estimation procedure, and 
consider both
the use of correlated probes as well as thermometry of correlated
environments. In particular, we investigate the use of two-qubit
quantum probes to estimate the temperature of different class 
of bosonic systems at equilibrium, either made of a single thermal 
bath interacting with the two qubits, or of two independent environments
having the same temperature, each one interacting locally with one of the
qubits. We compare performances with those obtained with a single 
qubit probe, and analyze in some details the interplay among the 
structure of the baths, the interaction time and the quantum 
correlations between the qubits in determining the overall 
thermometric precision.
\par
The paper is structured as follows. 
In section \ref{sec1} we  briefly review the main concepts of
 quantum estimation theory that will 
be used throughout the work.
 In Section \ref{sec2} we  present the physical model of a
 two-qubit system interacting with a bosonic thermal bath and 
 we analyze how its dynamics is affected by a common environment  or  two independent 
 and identical baths. 
 In Section \ref{sec3} we assess the role of entanglement in the precision of 
  the probing strategy  and then in 
   Section \ref{sec4} we  investigate the form of the POVM that guarantees optimal performances in 
 possible experimental implementations. Moreover we  test the robustness of the probes against initial perturbations  and we compare the performances of Bell/product states with 
 more general probes.
Section \ref{out} closes the paper with some %concluding
final remarks.
 %%%%%%%%%%%%%%
 %%%%%%%%%%%%%%
\section{Tools of quantum estimation theory}
\label{sec1}
In several sensing schemes, there is no direct access to the physical 
quantities of interest, which need to be evaluated by means of indirect measurements.  In turn, inferring the value of the quantity of interest 
by inspecting a set of data coming from the measurement of a different observable is precisely the goal of estimation theory. 
\par
Here we discuss how to optimally estimate the temperature $T$ 
of a (bosonic) thermal bath by performing a measurement on a quantum 
probe which is let interact with the bath, considered as its 
environment. 
The probe state is described by a density operator $\rho$ which, thanks 
to the interaction with the environment, becomes a function of the temperature of the bath 
$\rho \rightarrow \rho_T$. In other words, $T$, %is
the thermodynamical temperature of the bath, %but it is 
becomes just a parameter of the probe state having %and it has 
nothing to do with a possible probe temperature. %its temperature, which is not defined being its state out-of-equilibrium. 
This situation is at variance with classical thermometry, where 
the probe is let  interact with the sample until it reaches equilibrium.
Then the temperature read from the probe represents the thermodynamical temperature of both the sample {\em and the probe}. The larger set of 
available states and the inherent fragility of quantum states against decoherence %are making 
make quantum thermometry potentially more precise than
any classical protocol.
\par
In order to determine the parameter $T$, $M$ repeated 
measurements  of a probe observable $X$ are performed on identical preparations of the probe 
and the outcomes $\overline{x}=\lbrace x_1, x_2, \dots, x_M \rbrace$
 are then used to construct an estimator $\widehat{T}(\overline{x})$ of the temperature $T$. Hence, data  will be distributed around the mean value of the estimator according to the probability distribution $p(\overline{x}|T)=\prod_{k=0}^M p(x_k|T)$ (which is the conditional probability of obtaining the outcome $x_k$ when the parameter has value $T$) \cite{lqe09}:
\begin{equation}
\langle{T}\rangle=\int d\,\overline{x}\; p(\overline{x}|T)\, \widehat{T}(\overline{x}),
\end{equation}
with a certain variance $\sigma_T^2$, which characterizes the precision of the parameter estimation. At a classical level, for unbiased estimators, the variance of such distribution is bounded from below according by 
the Cram\`er-Rao inequality \cite{cramer}:
\begin{equation}
\sigma^2_T \geq \frac{1}{M F(T)},
\end{equation}
where $F(T)$ is the Fisher information for a single outcome,
\begin{equation}
F(T)=\int d\,x \frac{1}{p(x|T)}\left(\frac{\partial p(x|T)}{\partial T} \right)^2\,.
\end{equation}
In turn, $F(T)$ quantifies the amount of information carried by the random variable $x$ about the unknown parameter $T$. Upon optimising 
over all the possible quantum measurements one obtains the quantum Cram\`er-Rao bound
\begin{equation}
\sigma^2_T \geq \frac{1}{M F(T)}\geq \frac{1}{M H(T)},
\end{equation}
where $H(T)$ is the quantum Fisher information (QFI from now on). 
One of its explicit forms is:
\begin{equation}
\label{qfisher}
H(T)=2 \sum_{m,n}\frac{|\bra{\psi_m} \partial_T \rho_T \ket{ \psi_n}|^2}{\lambda_m+\lambda_n},
\end{equation}
with $\psi_n$ and $\lambda_n$ being respectively the 
$T$-dependent eigenvectors and eigenvalues of $\rho_T$.
Clearly, the optimal measurement is the one maximizing 
the QFI, while the the optimal estimator is that saturating
the inequality.
\par
The QFI can be written in the following form as well:
\begin{equation}
H(T)=\tr (\rho_T L^2_T)\,,
\end{equation}
where the symmetric logarithmic derivative (SLD) is given by
\begin{equation}
\label{sld}
L_T=2 \sum_{m,n} \frac{\bra{\psi_m} \partial_T \rho_T \ket{ \psi_n}}{\lambda_m+\lambda_n} \ketbra{\psi_m}{\psi_n}.
\end{equation}
This object is particularly useful since it can be proven that 
the optimal POVM is the spectral measure of the SLD. 
Related to the QFI, there is another figure of merit that gives a  quantification of the precision 
of the estimation. This is the quantum signal-to-noise ratio (QSNR), expressed as:
\begin{equation}
R(T) = T^2 H(T).
\label{qsnr}
\end{equation}
The expression of the QSNR in Eq. \eqref{qsnr} is derived from the ratio between the 
parameter  $T^2$ and the (single-measure) variance of the estimator. 
In this way,  a large value for $R$
means that the estimator  has a small relative error, i.e.  the error  is small  compared
to the value of the parameter to be estimated. 
The QFI, QSNR and the 
SLD are indeed the main tools that will be used in the  following.
\section{Physical model}
\label{sec2}
The single-qubit case has already been described in previous works, 
dealing with the \emph{purely dephasing} bath, related both to the estimation of temperature or other bath parameters \cite{benedetti18,razavian19,razavian2019,salary19}.
Here we focus, under the same conditions, on the two-qubit probe scenario, which allows us to explore the role 
of quantum correlations and the number of qubits in inferring the temperature.
We assume that initially the global state is separable
$\rho (0) = \rho_S (0) \otimes \rho_B (0)$,
where $\rho_B(0)$ is a thermal  state of the environment at inverse temperature $\beta$
and  characterized by the spectral
density:
\begin{equation}
J_s(\omega,\Omega)=\frac{\omega^{s}}{\Omega^{s-1}} e^{-{\omega}/{\Omega}},
\end{equation}
where $\Omega$ is the cutoff frequency  and $s$ is a Ohmicity parameter which distinguishes between sub-Ohmic ($s<1$), Ohmic ($s=1$) and super-Ohmic ($s>1$) regime.
The probe is composed of two qubits, which can be in a product or entangled initial state.
Moreover, the two qubits can be embedded in a common environment or in independent local baths.
%%%%%%%
%%%%%%%%%
\subsection{Common bath}
The Hamiltonian of two qubits interacting with a same bosonic environment can be written as:
\begin{equation}
\mathcal{H}= \sum_{j=1}^{2} \dfrac{\omega_j }{2}\sigma_3 ^{(j)} + \sum_{k=0}^{\infty} \omega_k b_k^\dag b_k  +\sum_{j=1}^2 \sum_{k=0}^{\infty}  \sigma_3 ^{(j)} (g_k b_k ^\dag + g_k^\ast b_k),
\end{equation}
where the index $j$ labels the qubits and the index $k$ labels the modes 
of the bath. $\sigma_3 ^{(j)} $ is the third Pauli matrix and $b(b^{\dagger})$ 
are the annihilation (creation) bosonic operators.
Moving to the continuum  $\sum_k  g_k\rightarrow 
 \int d\omega \,J_s(\omega,\Omega) (2 \,|g(\omega)|)^{-2}$ 
and letting the compound system probe plus bath evolve up to rescaled dimensionless time $\tau=\Omega t$ according to the above 
Hamiltonian, after tracing over the bath's degrees of freedom, one finds that the dynamics of the 
 two-qubit reduced density matrix can be expressed as~\cite{reina02}:
\begin{equation}
\rho_{\text{\tiny CB}} (\tau,T)=\mathcal{V}(\tau, T)\circ \mathcal{R}(\tau) \circ \rho, %\rho_ {\text{\tiny CB}}(0,T),
\label{cbmap}
\end{equation}
where $\mathcal{V}(\tau,T)$ is given by:
\begin{equation}
\mathcal{V}(\tau,T)=\begin{pmatrix}
1 & e^{-\Gamma_s(\tau,T)} & e^{-\Gamma_s(\tau,T)} & e^{-4 \Gamma_s(\tau,T)} \\
e^{-\Gamma_s(\tau,T)} & 1 & 1 & e^{-\Gamma_s(\tau,T)} \\
e^{-\Gamma_s(\tau,T)} & 1 & 1 & e^{-\Gamma_s(\tau,T)} \\
e^{-4\Gamma_s(\tau,T)} &e^{-\Gamma_s(\tau,T)} & e^{-\Gamma_s(\tau,T)} & 1
\end{pmatrix}
 \label{mapC}
\end{equation}
and
\begin{equation}
\label{rotation}
\mathcal{R}(\tau)=\begin{pmatrix}1 & e^{2if(\tau)} & e^{2if(\tau)} & 1 \\
e^{-2if(\tau)} & 1 & 1 & e^{-2if(\tau)} \\
e^{-2if(\tau)}& 1 & 1 & e^{-2if(\tau)} \\
1 & e^{2if(\tau)} & e^{2if(\tau)} & 1
\end{pmatrix},
\end{equation}
and $\circ$ is the Hadamard (entrywise product), %$\tau=\Omega t$ 
%is a dimensionless time, 
while %$\rho_ {\text{\tiny CB}}(0)$
$\rho$ is the %common-bath
initial state of the qubits.
The decoherence function depends on time and the  dimensionless temperature 
$T=(\Omega \beta)^{-1}$. By expressing also $\omega$ in unit of $\Omega$, i.e.
$\omega \rightarrow \omega/\Omega$, we may write
\begin{align}
\Gamma_s (\tau,T)&=\int_0^\infty \! \! d\omega\,  e^{-\omega} \,
\frac{ 1- \cos \omega \tau }{\omega^{2-s}} 
\coth\left(\frac{\omega}{2 T} \right)\,,
\label{gamma}
\end{align}
whose analytic expression can be found in \cite{razavian2019}. 
The function $f(\tau)$ is instead a temperature-independent quantity
\begin{equation}
\label{rotationcontinuum}
f(\tau)=\frac{1}{2} \int_0^\infty \!\!\! d\omega\,J_s(\omega)\,\frac{\omega \tau- \sin \omega\tau}{\omega^2}\,.
\end{equation}
The  matrix $\mathcal{V}(\tau,T)$ is dominant at short times and the matrix $\mathcal{R}(\tau)$ is 
able to create entanglement between the qubits.
%%%%%%%%%%%%%%%%%%%
%%%%%%%%%%%%%%%%%%%
\subsection{Independent baths}
In the case the two qubits interact with  identical but independent local environments,
the Hamiltonian  takes the form:
\begin{align}
\mathcal{H}\!= \! \sum_{j=1}^{2} \! \left[\dfrac{\omega_j }{2}\sigma_{\!3} ^{\!\tiny (j)}  \!\! +\sum_{k=0}^{\infty} \omega_k b_k^{\!(j)\dag} b_k^{\!\tiny(j)} \!\! +\sigma_3 ^{(j)} \! \sum_{k=0}^{\infty} \!\left(g_k b_k ^{\! (j)\dag} \!\!+ g_k^\ast b_k^{\!(j)}\!\right)\! \right]
\nonumber
\end{align}
Let the initial density matrix of the bath be factorized
$\rho_b(0)= \rho_B^{(1)} \otimes \rho_B^{(2)}$ 
where  $\rho_B^{(1,2)}$  are thermal Gibbs states at dimensionless temperature $T$.
In this case, the evolution of the probe %system  
is given by:
\begin{equation}
\rho_{\text{\tiny LB}}(\tau,T)=\mathcal{W}(\tau,T) \circ \rho,%_{\text{\tiny LB}}(0,T),
\end{equation}
where $\mathcal{W}(\tau,T)$ is the tensor product of two  single qubit dephasing maps
\begin{equation}
\mathcal{W}(\tau,T)\!=\! \begin{pmatrix}
1&e^{-\Gamma_s(\tau,T)}&e^{-\Gamma_s(\tau,T)}&e^{-2\Gamma_s(\tau,T)}\\
e^{-\Gamma_s(\tau,T)}& 1&e^{-2\Gamma_s(\tau,T)}&e^{-\Gamma_s(\tau,T)}\\
e^{-\Gamma_s(\tau,T)}&e^{-2\Gamma_s(\tau,T)}&1&e^{-\Gamma_s(\tau,T)}\\
e^{-2\Gamma_s(\tau,T)}&e^{-\Gamma_s(\tau,T)}&e^{-\Gamma_s(\tau,T)}&1
\end{pmatrix}. \label{mapI}
\end{equation}
 
%%%%%%%%%%%%%%%%%%%%%%%%
\section{Two-qubit thermometry}
\label{sec3}
We now compare the performances of two-qubit quantum probes for the temperature by analyzing the 
behaviour of the QFI. We consider two different initial states for the qubits, i.e.
 initially entangled and initially separable probes and we study the role of 
 local against common baths.
%%%%%%%%%%%%%%%%%%%%%%%%
%%%%%%%%%%%%%%%%%%%%%%%%
\subsection{Entangled qubits}\label{res1}
We start by analyzing the performances of quantum probes initially prepared in one of the Bell state:
$
\ket{\varphi^{\pm}}=\frac{1}{\sqrt{2}}(\ket{00} \pm \ket{11})$ and $\ket{\psi^{\pm}}=\frac{1}{\sqrt{2}}(\ket{01} \pm \ket{10})$.
We firstly consider the scenario with a common environment: 
the Bell states $\ket{\psi^{\pm}}$ are not changed by the map \eqref{cbmap},
while $\ket{\varphi^{\pm}}$ are affected by the environment. Indeed, for both states $\ket{\varphi^{\pm}}$ 
the QFI is:
\begin{equation}
H_{\text{\tiny CB}}^{\text{\tiny ent}}(\tau,T,s)=\frac{ \big[4\,\partial_{\text{\tiny {\it T}}} \, \Gamma_s(\tau,T) \big]^2}{e^{8\,\Gamma_s(\tau,T)}-1}.
\label{entcb}
\end{equation}
Notice that this quantity does not depend on $f(\tau)$.
On the other hand, when the interaction of the probe with two identical and independent local baths is considered, all four Bell states lead to the same QFI:
\begin{equation}
H^{\text{\tiny ent}}_{\text{\tiny LB}}(\tau,T,s)=\frac{ \left[2\, \partial_{\text{\tiny {\it T}}} \,\Gamma_s(\tau,T)\right]^2}{e^{4\Gamma_s(\tau,T)}-1}.
\label{entlb}
\end{equation}
A numerical evaluation of the QFI shows that the  Bell states 
are optimal within the subset of states 
\begin{eqnarray}
\label{tool0a}
\ket{\Phi_{\alpha}}&=& \cos \alpha \ket{00} + \sin \alpha \ket{11}\\
\label{tool0b}
\ket{\Psi_{\alpha}}&=&\cos \alpha \ket{01} + \sin \alpha \ket{10}\ ,
\end{eqnarray}
 with $\alpha\in [0,2\pi]$ a real parameter.
%%%%%%%%%%%%%%%%%%%%%%%%%
%%%%%%%%%%%%%%%%%%%%%%%%%
\subsection{Separable qubits}\label{res2}
Let us consider two qubits initially prepared in the product state
$ \ket{++}$, where $\ket{+}=\frac{1}{\sqrt{2}}(\ket{0}+\ket{1})$,
which is a particular instance of the one-parameter family of states 
\begin{equation}
\label{tool1}
\ket{\psi_\alpha\psi_\beta}:=\ket{\psi_{\alpha}}\otimes \ket{\psi_{\beta}}\ ,\quad 
\ket{\psi_{\alpha}}=\cos\alpha\ket{0}+\sin\alpha\ket{1}.
\end{equation}
In the common-bath case, the QFI has the expression:
\begin{align}
&H^{\text{\tiny sep}}_{\text{\tiny CB}}(\tau,T,s)=\nonumber \\& 
\frac{4(2-e^{2\Gamma_s(\tau,T)}+2e^{4\Gamma_{\!s}(\tau,T)}+e^{6\Gamma_s(\tau,T)})[\partial_{\text{\tiny T}}\Gamma_s(\tau,T)]^2}{3\,e^{8\,\Gamma_s(\tau,T)}-2\,e^{4\,\Gamma_s(\tau,T)}-1}.
\label{sepcb}
\end{align}
Again, the quantity $f(\tau)$ does not play a role in the estimation of the temperature.
In the case of local independent baths, instead, the QFI reads:
\begin{equation}
H^{\text{\tiny sep}}_{\text{\tiny LB}}(\tau,T,s)=\frac{2[\partial_{\text{\tiny T}} \Gamma_s(\tau,T)]^2}{e^{2\,\Gamma_s(\tau,T)}-1}.
\label{seplb}
\end{equation}
As expected  it is twice the QFI of a single-qubit probe \cite{razavian2019}.
Both results \eqref{sepcb} and  \eqref{seplb} can be obtained by initializing the 
probe also in one of the states $\ket{--}$, 
$\ket{+-}$ and $\ket{-+}$.
\par
\begin{figure}[t]
\centering
\includegraphics[width=0.99\columnwidth]{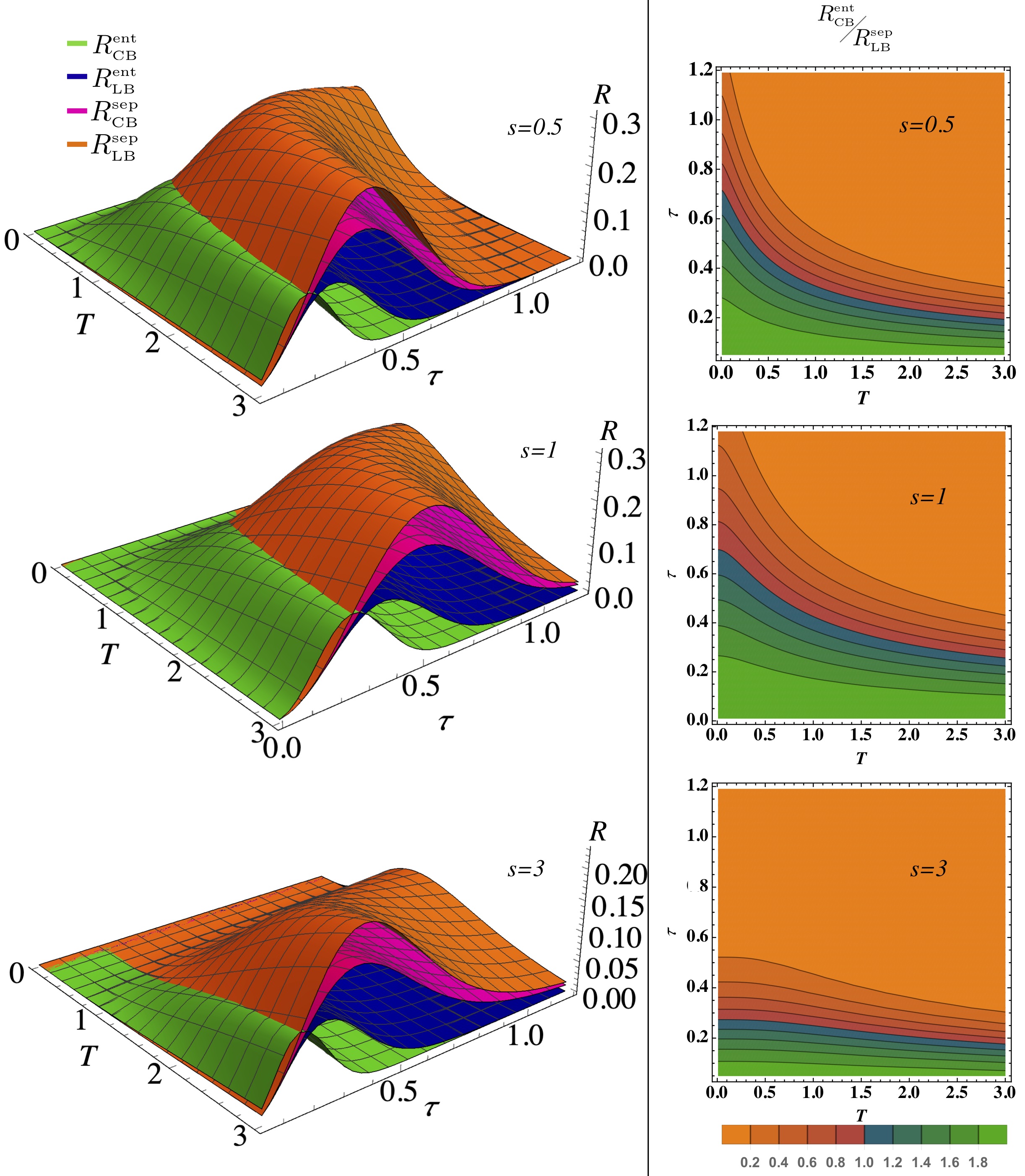}
\caption{
Left column: behaviour of the QSNR for three different values of $s=0.5,\,1,\,3$, as a function of 
dimensionless time $\tau$ and temperature $T$.
The four different colors correspond to the possible scenarios analyzed in this
work, i.e. common/local bath and entangled/separable initial state.
Right column: Ratio $R_{\text{\tiny CB}}^{\text{\tiny ent}}/R^{\text{\tiny sep}}_{\text{\tiny LB}}$ 
for the same three values of $s$.
}
\label{comp01}
\end{figure}

Let us now compare the behaviour of the quantum signal-to-noise ratio (QSNR) given in (\ref{qsnr}), that for sake of simplicity we shall denote by $R$ in the following, for different initial states and bath parameters.
In figure Fig.~\ref{comp01}, we show the behaviour of $R$ for different values of the Ohmicity parameter as a function of the dimensionless time and temperature and for the four 
possible scenarios described in section \ref{res1}
and \ref{res2}.
The evaluation of  the QSNR can only be performed numerically, through the integration of the decoherence function \eqref{gamma}.
From Fig. \ref{comp01}, a universal behavior emerges which is independent on the Ohmicity parameter. Indeed,  
we found two
optimal scenarios  that give the maximum of the QSNR $R$. 
For short time intervals (namely $\tau \ll 1$), the optimal strategy to estimate the temperature is to 
employ an entangled probe in a common bath. For longer times, the estimation
is more precise if a separable initial state  is left evolve in an environment consisting of local independent baths.
It follows that, if time is a resource that we need to use parsimoniously, the best option is to use
an entangled probe interacting with a common environment. On the contrary,
if we can wait for longer times, the (absolute) best strategy is sending
sequentially  two qubits prepared in a $\ket{+}$ state through the bath, i.e.
repeating twice the single-qubit scheme \cite{razavian2019}.
In Fig. \ref{comp01}, we  also analyze the ratio 
$R_{\text{\tiny CB}}^{\text{\tiny ent}}/R^{\text{\tiny sep}}_{\text{\tiny LB}}$
to emphasize the  optimal procedures in the $(\tau,T)$-space.
As we see from the graphs, the Ohmicity parameter $s$  only affects the qualitative behaviors of the ratio, and 
the two different estimation strategies are clearly displayed.
In order to better understand why the change  from one optimal strategy to the other occurs, 
we examine the  behaviour of the quantum Fisher information $H$ for small values of $\tau$:
\begin{equation}
\label{short}
H(T,\tau,s) \simeq \tau^2\left\{ \begin{array}{ccl}
\frac1{\gamma(T,s)} \left[\partial_T \gamma(T,s)\right]^2 && \hbox{for}\, H_{\text{\tiny CB}}^{\text{\tiny ent}}\\
 & & \\
\frac1{2\,\gamma(T,s)} \left[\partial_T \gamma(T,s)\right]^2 && \hbox{other cases},
\end{array}\right.
\end{equation}
where
\begin{equation}
\Gamma_s(T,\tau) \simeq \frac{\tau^2}{2} \int_0^\infty dx \; x^s e^{-x} \coth{\frac{x}{2T}}\equiv \frac{\tau^2}{2}\vspace{2cm} \gamma(T,s)
\vspace{-2cm}
\end{equation}
at short times.
So, $H_{\text{\tiny CB}}^{\text{\tiny ent}}$ outperforms by a factor $2$  the other strategies in the $\tau\ll1$ regime. 
\\
\begin{figure}[t!]
\centering
\includegraphics[width=0.99\columnwidth]{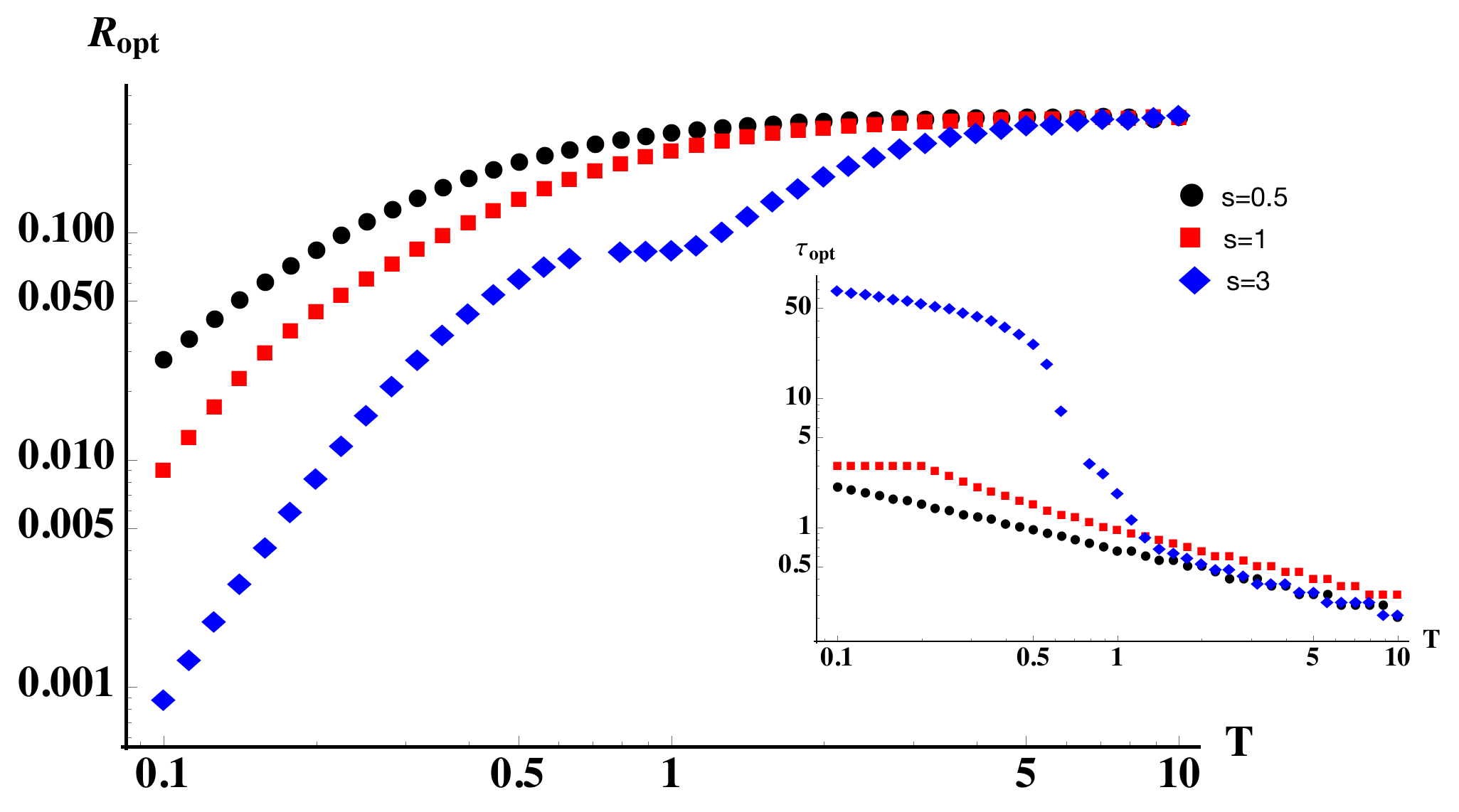}
\caption{Maximum value of the QSNR in time as a function of the temperature for different values of $s$.
This maximum is obtained by  separable probes in local baths, as shown in Fig. \ref{comp01}.
In the inset, optimal time, i.e. time at which $R$ reaches the maximum value, as a function of $T$.}
\label{optimal}
\end{figure}

In Fig. \ref{optimal}, we show the maximum value of the QSNR, called $R_{\text{\tiny opt}}$,
as a function of the temperature and for different values of the Ohmicitiy parameter $s$.
Notice that this maximum is obtained in the case 
where two separable qubits are embedded in local baths. 
We notice that, as the Ohmicity parameter  $s$ grows, the maximal values of the QSRN decrease for a fixed value of the temperature, except for $T$ becoming large and the dependency on $s$ disappears.
The time corresponding to the  maximum of $R$, $R_{opt}$,   is called optimal time $\tau_{\text{\tiny{opt}}}$; notice that, especially for low temperatures, it strongly depends on the Ohmicity parameter. Furthermore, Fig. \ref{optimal} shows  agreement with the 
behaviour found in the case of single qubit thermometry \cite{razavian2019}. This is expected since the maximum of the 
QSNR is obtained employing two qubits in a product states that evolve in local baths. 
This procedure is the same as sending twice a single qubit into the quantum environment.
\section{Implementations}
\label{sec4}
In this section we analyze possible ways to experimentally implement the optimal strategies. 
We focus on  
entangled qubits  in 
a common bath at short times, and  separable qubits in independent baths for
longer interaction times. 
By calculating the symmetric logarithmic derivative, we  derive the optimal POVM to be performed on the qubits 
in order to infer the value of temperature in a bosonic environment.
\subsection{SLD for Bell states in a common bath}
\begin{figure}[!t]
\centering
\includegraphics[width=0.75\columnwidth]{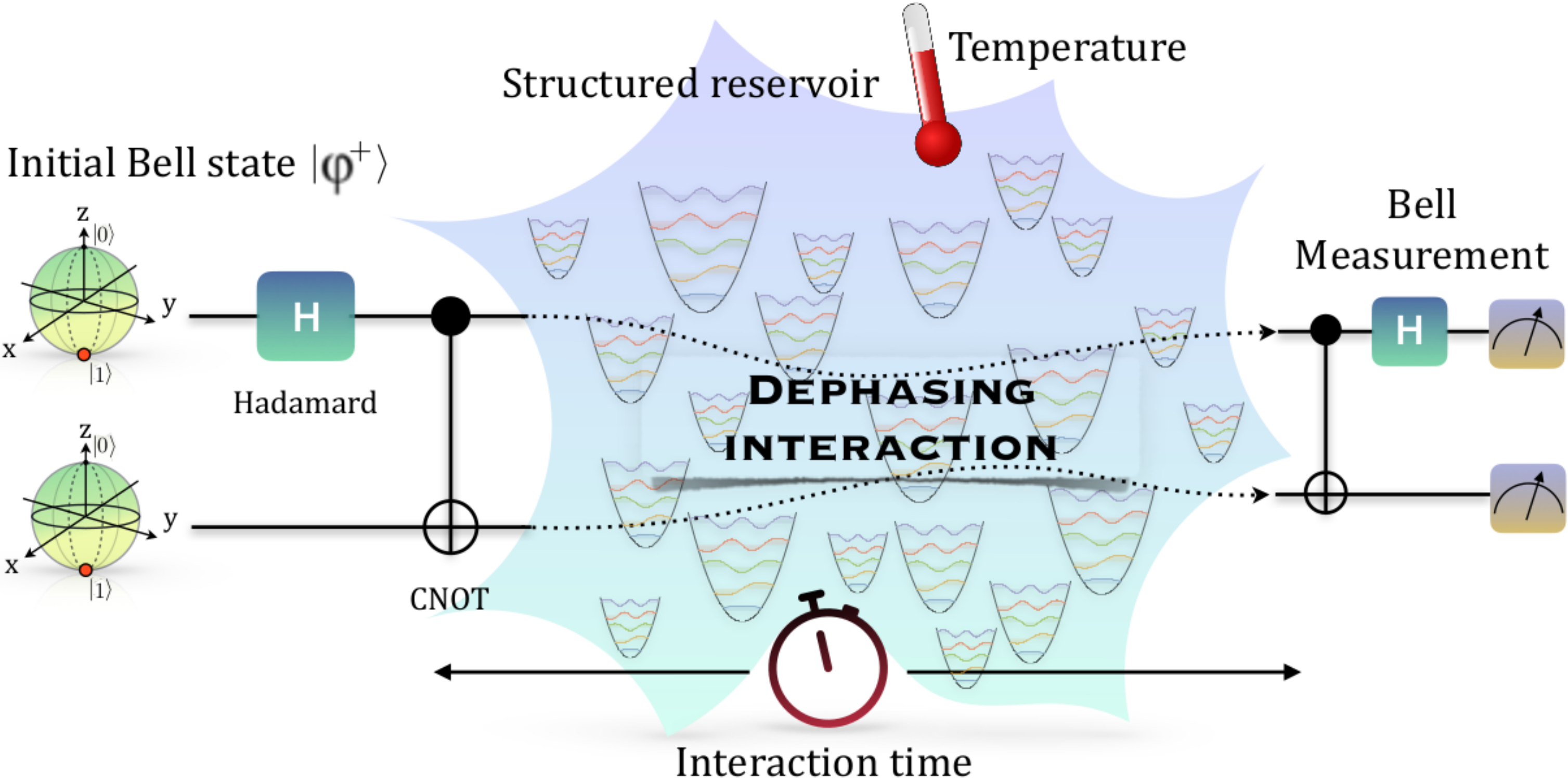}
\caption{General scheme for state $\ket{\varphi^+}$ in the same bath. The 
optimal measurement consists in a Bell measurement along two Bell states.}\label{im:povment}
\end{figure}

We begin by considering the scenario of the optimal Bell states $\ket{\varphi^\pm}$ in a common bath.
The SLD is derived using equation  (\ref{sld}):
%\begin{align}
%L_e^{C}=b(\tau,T,s)(\ket{\varphi^+}\bra{\varphi^+}+\ket{\varphi^-}\bra{\varphi^-}),
%\end{align}
%where 
%\begin{equation}
%b(\tau,T,s)=-\frac{4 \partial_T \Gamma_s(\tau,T)}{e^{8 \Gamma_s(\tau,T)}+1}.
%\end{equation}
\begin{equation}
L_{\text{\tiny CB}}^{\text{\tiny ent}}=a_{-}\ketbra{\varphi^{-}}{\varphi^{-}}-a_{+}\ketbra{\varphi^{+}}{\varphi^{+}},
\end{equation}
where
\begin{equation}
a_{\pm}=\frac{4\,\partial_{\text{\tiny T}}\Gamma_{s}(\tau,T)}{e^{4\Gamma_{s}(\tau,T)}\pm1}.
\end{equation}
The coefficients of the decomposition depend on the temperature, time and the Ohmicity, 
but  the projectors are temperature- and Ohmicity-independent.
The optimal protocol thus requires to experimentally discriminate  the Bell states $\ket{\varphi^\pm}$. 
This general scheme is sketched in Fig.\ref{im:povment}.

\begin{figure}[!t]
\centering
\includegraphics[width=0.75\columnwidth]{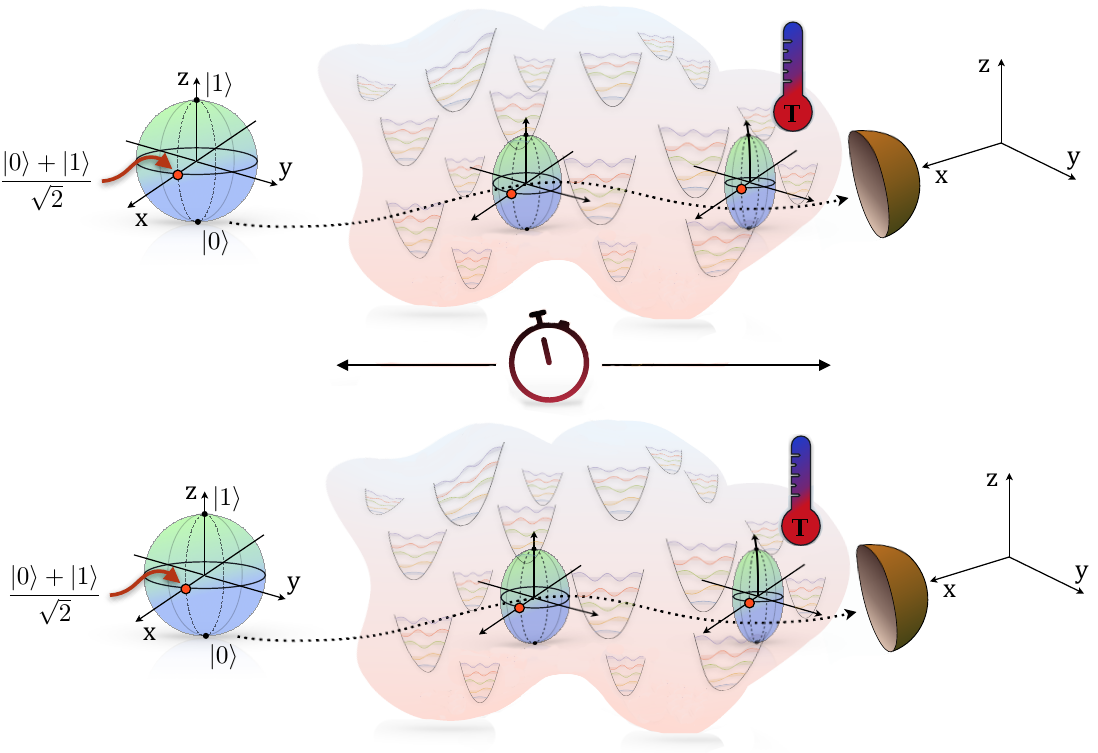}
\caption{ General scheme of two qubits in the $\ket{++}$  state interacting with independent 
baths and being measured separately along $\sigma_1$.}\label{im:povmpro}
\end{figure}
%%%%%%
\subsection{SLD for $\ket{++}$ states in local baths}
Regarding the separable state $\ket{++}$ coupled with two independent local baths, 
  the SLD has the expression:
\begin{align}
&L_{\text{\tiny LB}}^{\text{\tiny sep}}=b_{-}\Pi_{-}
-b_{+}\Pi_{+}+c\left(\ketbra{\varphi^-}{\varphi^-}+\ketbra{\psi^-}{\psi-}\right)
\label{sldpp}
\end{align}
where
\begin{align}
b_{\pm}=\frac{2\,\partial_{\text{\tiny T}}\Gamma_{\!s}(T,\tau)}{e^{\Gamma_{\!s}(T,\tau)}\pm1}\qquad
c=\frac{2\,\partial_{\text{\tiny T}}\Gamma_{\!s}(T,\tau)}{e^{2\Gamma_{\!s}(T,\tau)}-1}
\end{align}
and $\Pi_{\pm}=\ketbra{\pm \pm}{\pm \pm}$.
Although the presence of Bell states in Eq. \eqref{sldpp} may led to think that global measurements on both qubits are needed in order to discriminate the various projections, actually this is not the case: the eigenstates of $L_{\text{\tiny LB}}^{\text{\tiny sep}}$ are eigenvectors of  $\sigma_1 \otimes \sigma_1$, an observable that can be accessed locally, namely by separate measurements on the two qubits. 
This scenario is sketched in Fig.\ref{im:povmpro}.
%%%%
\subsection{Robustness}
\label{subsectionrob}
We want to check the robustness of the optimal states, namely  we want to quantify how 
 the maximal QFI is affected by  small deviations $\delta$ from the optimal states.
 We start by focusing on the two optimal scenarios. 
As already mentioned, sending two qubits  in a product state in independent local environments corresponds 
to repeating the same experiment twice; therefore we can write the QFI for a generic initial state $\ket{\psi_\alpha \psi_\beta}$ defined in~\eqref{tool1} as the sum of the QFI of the two single qubits:
\begin{align}
\label{qfigsep}
H^{\alpha \beta}_{\text{\tiny LB}}(\tau,T)=%&H_{ \alpha} (\tau,T)+ H_{\beta}(\tau,T)= \\
\left[\sin^2(2 \alpha)+\sin^2(2 \beta) \right] \frac{ (\partial_T \Gamma_s(\tau,T))^2}{e^{2\Gamma_s(\tau,T)}-1}.
\end{align}
We now consider the  perturbed  initial state  $\ket{\psi_{\alpha+\delta_{\alpha}} \psi_{\beta+\delta_{\beta}}}$ around the optimal choice ($\alpha=\beta=\pi/4$),
we find the perturbed QFI:
\begin{equation}
H^{\delta_{\alpha} \delta_{\beta}}_{\text{\tiny LB}}(\tau,T)= H^{\alpha \beta}_{\text{\tiny LB}}(\tau,T)\left[1- 2\left(\delta_{\alpha}^2 + \delta_{\beta}^2\right)\right].
\end{equation}
On the other side,  the general entangled state $\ket{\Phi_{\alpha}}$ defined in~\eqref{tool0a} in a common bath  yields
\begin{align}
\label{qfigent}
&H^{\alpha}_{\text{\tiny CB}}(\tau,T)=\frac{16 \sin^2(2\alpha)\,[\partial_{\text{\tiny T}}\rho_{\text{\tiny T}}]^2}{e^{8\Gamma_{\!s}(\tau,T)}-1}.
\end{align}
If we perturb the state $\ket{\Phi_{\alpha}}$ around $\alpha=\pi/4$, we find the following QFI:
\begin{equation}
H^{\delta}_{\text{\tiny CB}}(\tau,T) \simeq H^{\text{\tiny ent}}_{\text{\tiny CB}}(\tau,T) \left(1- 4\delta^2\right).
\end{equation}
In both cases  the deviation from the optimal QFI are at second order on the magnitude of the perturbation, i.e. the QFI obtained from  both $\ket{\varphi^+}$ and $\ket{++}$ 
is robust with respect to the probe preparation.
%
%%%%%%%%%%%%%%%%%%%%%%%%%%
\subsection{Performances of other probes}
\label{subsecother}
\begin{figure}[!t]
\centering
\includegraphics[width=0.98\columnwidth]{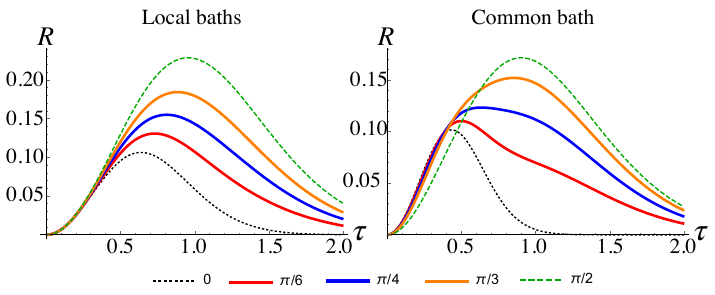}
\caption{QSNR $R$ as a function of time for $s=1$, $T=1$ and  different values of the parameter $\alpha$ in Eq. \eqref{superS}.  
Both local- and common-bath scenarios are represented.}
\label{im:sovralfa}
\end{figure}

So far  we maximized the QFI within specific subsets of state,  $\ket{\Phi_{\alpha}}$ and  $\ket{\psi_{\alpha}\psi_{\beta}}$
 for  entangled and separable probes and
 we found that, within these families, the states $\ket{\varphi^+}$ and $\ket{++}$ are 
  optimal  regardless of $\tau$ and $T$.  
 We now  consider the normalized superposition:
\begin{equation}
\ket{S_{\alpha}}=C\left(\cos{\alpha}\ket{\varphi^+}+\sin\alpha\ket{++}\right),
\label{superS}
\end{equation}
where $C$ is a normalization constant.
In Fig.\,\eqref{im:sovralfa} we  compare the behaviour of the QSNR  for $s=1$ and $T=1$ for different values of the parameter $\alpha$ in Eq. \eqref{superS}. 
%{Since we fixed the parameter $T=1$, then the QSNR coincides with the QFI, see Eq. \eqref{qsnr}, in this specific case.}
Both the local and global scenarios are analyzed. 
 The  largest value of the QSNR is obtained  for $\alpha=\pi/2$, corresponding to initially separable qubits.
 At shorter times, the $\alpha=0$ case, i.e. entangled qubits, is the best probe, but only in the common-bath case.
 In the common-environment scenario, moreover, there is also a small time-interval
  where other probes perform better,
 while this behaviour is never seen in the local-bath case.
\begin{figure}[!t]
\centering
\includegraphics[width=0.98\columnwidth]{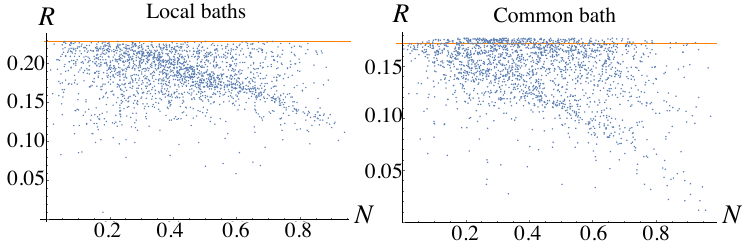}
\caption{Maximum value of the QSNR $R$ as a function of the entanglement $N$ of the initial state of the probes in a time interval $\tau\in [0.2]$. 
2000 pure states are generated randomly, both for the local- and common-bath scenario. 
We set $s=1$ and $T=1$. The orange line is a guide for the eye for the maximum value of the QSNR reached by the probe $\ket{++}$.}
\label{im:random}
\end{figure} 

In order to investigate the role of other probes, we compare the QSNR
 for randomly-generated pure states.
In Fig.\,\eqref{im:random}, we display the behaviour of the maximum of the QSNR in a time interval $\tau\in [0,2]$ as a function of
the entanglement of initial state of the probe. We generated 2000 points, representing different initial pure states with a different level of entanglement.
Entanglement is evaluated through the negativity  $N$ \cite{neg}.
We see that, in the local-bath scenario, the separable state $\ket{++}$ achieves indeed the maximum of the QSNR, marked by the orange line.
All other generated states have a QSNR that stays below this value.
However, this is not the case in the common-bath scenario, where there exist entangled probes that perform better than the $\ket{++}$ state and that cannot be
described through Eq.\,\eqref{superS}.  These states, in fact, outperform the product state in estimating the temperature of the
bath, as shown on the right plot of Fig.\,\eqref{im:random}.
%%%%%%%%%%%%%%%%%%%%%%%%%%%%%%%%
%%%%%%%%%%%%%%%%%%%%%%%%%%%%%%%%
%
\section{Conclusions}\label{out}
The aim of thermometry is  to estimate the temperature of a certain 
object without altering its properties, temperature included.  When 
the object is a quantum system, the idea of finding a non-invasive 
probe becomes a necessity, in order to avoid decoherence. In this paper,
we have addressed optimal quantum thermometry for bosonic 
thermal baths by means of two-qubits quantum probes. 
We have analyzed  the case of two independent baths, each 
interacting locally with a qubit, and the case of a common
reservoir acting upon the two qubits. Moreover, we have 
considered two families of initial states for the probes: Bell 
states and product states of the form $\ket{\sigma,\eta}$ 
where $\sigma={\pm}$ and $\eta=\pm$. In particular, we have 
analyzed the behaviour of the QFI and the QSNR as a function of
time, temperature and ohimicity parameter. Our results show that
at short times the best way to probe the temperature of the bath 
is to employ maximally entangled qubit interacting with the same 
global bath, independently from the  Ohmicity of the environment. 
However, if time is not considered a resource, we found that 
the  best estimate is obtained upon using two qubits in a 
product states, each one subject to a local environment, i.e. 
the best strategy is to repeat twice a single-qubit probe 
measurement \cite{razavian19,razavian2019}. 
Notice that the notions of {\em same bath} or {\em two independent} 
replicas of the same bath do not require multiple physical systems to be implemented in practice. In a realistic environment, the two situations correspond respectively to the two qubits propagating one close to each other, in order to sense the same portion of the environment, or far away, such that possible spatial correlations of the bath may be neglected. We have analysed possible implementations of the optimal estimation schemes discussed above and we found that the entangled case boils down to a Bell measurement over the entangled quits, while the case of factorized probe may be implemented 
by a local measurement performed separately on the two qubits.
We have analyzed the behaviour of the estimation precision against 
perturbations in the preparation of the initial state and
have shown the they contribute to decrease the QFI 
with second order corrections in the 
perturbations, i.e. our scheme is robust.
\par
We have also compared the performances of the Bell and product 
states with other probes. Our  intent was to analyze whether a 
combination of the  Bell and separable states could improve the 
precision of the estimation. We showed that the absolute maximum 
of the QFI is indeed obtained for independent qubits in a product state. 
However, there exists temporal regions where a superposition of the
two probes perform better, though only in the common-bath scenario.
In order to gain more insight into the  role of other probes, we have 
randomly generated pure states and analyzed the maximum value of 
their QFI versus the  entanglement of the generated state. We found 
numerical evidence that the state $\ket{++}$ is the optimal one in 
the case of local environments, while for the common bath there 
exist initial entangled states of the probe that outperform the product state 
in a certain temporal region that depends upon the temperature and 
the Ohmicity parameter. The corresponding optimal measurement, however, 
do not correspond to an easy implementable one and thus it may be 
challenging to achieve the corresponding enhancement in a practical
scenario. 
\par
Summarizing our results, we have that entanglement 
improve thermometry at short times whereas, if the interaction 
time  is not constrained, coherence rather than entanglement, 
is the key resource to improve precision. We also emphasise
that our scheme for quantum thermometry is based on pure dephasing 
and it does not involve energy exchanges between the probes and 
the bath. The corresponding measuring protocols are thus 
inherently non invasive, and they may be of value 
whenever a direct inspection threatens 
to destroy the sample, or to perturb the temperature itself, as it 
would happen in the case of very cold samples. 
\acknowledgements
MGAP is member of INdAM-GNFM. CB, MB and MGAP thanks S. Razavian, 
I. Amelio and F. Grasselli for useful discussions. 
This work has been partially supported by JSPS through 
FY2017 program (grant S17118).

\end{document}